\documentclass[sn-mathphys,Numbered]{sn-jnl}

\usepackage{graphicx}%
\usepackage{multirow}%
\usepackage{amsmath,amssymb,amsfonts}%
\usepackage{amsthm}%
\usepackage{mathrsfs}%
\usepackage[title]{appendix}%
\usepackage{xcolor}%
\usepackage{textcomp}%
\usepackage{manyfoot}%
\usepackage{booktabs}%
\usepackage[ruled,vlined]{algorithm2e}
\usepackage{graphicx}
\usepackage{float}
\usepackage{subfigure}
\usepackage{tabularx}
\usepackage{algpseudocode}%
\usepackage{listings}%

\theoremstyle{thmstyleone}%
\theoremstyle{thmstyletwo}%

\theoremstyle{thmstylethree}%

\begin{document}

\title[Article Title]{Knowledge Graph Enhanced Intelligent Tutoring System Based on Exercise Representativeness and Informativeness}


\author[1]{\fnm{Linqing} \sur{Li}}\email{a847820455@gmail.com}

\author*[1,2]{\fnm{Zhifeng} \sur{Wang}}\email{zfwang@ccnu.edu.cn}

\affil[1]{\orgdiv{CCNU Wollongong Joint Institute},  \orgname{Central China Normal University}, \orgaddress{\street{Luoyu Road}, \city{Wuhan}, \postcode{430079}, \country{China}}}

\affil*[2]{\orgdiv{Faculty of Artificial Intelligence in Education}, \orgname{Central China Normal University}, \orgaddress{\street{Luoyu Road}, \city{Wuhan}, \postcode{430079}, \country{China}}}


\abstract{In the realm of online tutoring intelligent systems, e-learners are exposed to a substantial volume of learning content. The extraction and organization of exercises and skills hold significant importance in establishing clear learning objectives and providing appropriate exercise recommendations. Presently, knowledge graph-based recommendation algorithms have garnered considerable attention among researchers. However, these algorithms solely consider knowledge graphs with single relationships and do not effectively model exercise-rich features, such as exercise representativeness and informativeness. Consequently, this paper proposes a framework, namely the Knowledge-Graph-Exercise Representativeness and Informativeness Framework, to address these two issues. The framework consists of four intricate components and a novel cognitive diagnosis model called the Neural Attentive cognitive diagnosis model. These components encompass the informativeness component, exercise representation component, knowledge importance component, and exercise representativeness component.
The informativeness component evaluates the informational value of each question and identifies the candidate question set ($Q_C$) that exhibits the highest exercise informativeness. Moreover, the exercise representation component utilizes a graph neural network to process student records. The output of the graph neural network serves as the input for exercise-level attention and skill-level attention, ultimately generating exercise embeddings and skill embeddings. Furthermore, the skill embeddings are employed as input for the knowledge importance component. This component transforms a one-dimensional knowledge graph into a multi-dimensional one through four class relations and calculates skill importance weights based on novelty and popularity. Subsequently, the exercise representativeness component incorporates exercise weight knowledge coverage to select questions from the candidate question set for the tested question set. Lastly, the cognitive diagnosis model leverages exercise representation and skill importance weights to predict student performance on the test set and estimate their knowledge state.
To evaluate the effectiveness of our selection strategy, extensive experiments were conducted on two publicly available educational datasets. The experimental results demonstrate that our framework can recommend appropriate exercises to students, leading to improved student performance.}

\keywords{multi-dimensional knowledge graph, graph neural network, exercise recommendation, cognitive diagnosis model}



\maketitle

\section{Introduction}\label{sec1}

Online education has emerged as a significant supplementary learning strategy for students \cite{sun2021design,cui2023survey}. Many students rely on online exercise recommendations to support their learning. With the vast amount of educational materials available, the challenge lies in recommending appropriate exercises for effective learning \cite{idris2009adaptive,ma2022exercise,kang2019personalized,xia2018personalized}. Exercises play a crucial role in personalized educational services by serving as a powerful tool to assess students' mastery of concepts. However, given the abundance of exercise resources, it is nearly impossible for students to complete them all within a limited time. Therefore, assisting students in finding suitable exercises becomes a significant problem. An exercise recommendation system has been proposed to address this issue by leveraging students' historical answer sequences \cite{wu2020exercise,ai2019concept,stout2021systematic}. 

Knowledge graphs (KGs), also known as cognitive maps, provide graphical representations where concepts or words are organized into nodes and connected by vectors representing relationships. The application of knowledge graphs to learn the order of skills has shown promising results (e.g., \cite{kamsa2016optimizing,jia2018representation,karim2022question,pellegrino2021automatic,martin2021personalization}). The arrangement of concepts in a knowledge graph significantly impacts learning ability \cite{sanderson2002learning,saxena2022sequence,wang2019exploring}.

Researchers have recognized that exercise and skill features in knowledge graphs greatly influence the quality of learning when recommending exercises \cite{romero2006data,lv2018utilizing,zhu2020study}. Various methods have been developed to learn the features of knowledge graphs and recommend appropriate exercises, resulting in improved student performance \cite{shi2020learning,bi2020quality,lv2021intelligent,zhao2019mathgraph}. These methods effectively explore skill and exercise features to enhance learning efficiency. Additionally, high-quality exercises contribute to learners' comprehension of the learning material. Consequently, the research community strives to create a high-quality exercise set to enhance e-learners' performance. Previous research \cite{zhu2018multi,kamsa2016optimizing} applied KGs to consider the dependencies of learning objects in exercise recommendations. However, these works only focused on basic relationships to establish links between KGs, without further investigating exercise features during the recommendation process. As a result, these methods fall short of meeting the requirements of modern e-learning.

This paper presents an innovative framework called Knowledge-Graph-Exercise Informativeness and Representativeness (KG-EIR) to address diverse learning needs based on KGs. To recommend exercises with high learning quality, the KG-EIR framework combines multidimensional KGs with exercise features to define the recommendation goal and enhance exercise quality. The KG-EIR framework consists of four innovative components and a novel cognitive diagnosis model called the Neural Attentive Cognitive Diagnosis model (NACD), which facilitates exercise recommendation to achieve the recommendation goal. The four components are the informativeness component, exercise representation component, knowledge importance component, and exercise representativeness component. The recommendation goal involves recommending exercises with high informativeness and representativeness.

Specifically, the informativeness component aims to select questions with high informativeness from the untested question set ($Q_U$) to the candidate question set ($Q_C$). The exercise representation component incorporates a graph neural network with two types of attention mechanisms to generate exercise and skill embeddings. The knowledge importance component utilizes an innovative knowledge points extraction algorithm that incorporates skill embeddings to extract knowledge points based on the multidimensional KG. Five skill features of these knowledge points are discussed to generate skill importance weights. Subsequently, the exercise representativeness component selects questions with high knowledge coverage from the candidate question set ($Q_C$) to the tested question set ($Q_T$) to achieve representativeness objectives. Finally, the NACD model predicts student performance on the tested question set and estimates their current knowledge state.

The main contributions of this paper can be summarized as follows:
\begin{enumerate}
	\item We propose a novel exercise recommendation method, KG-EIR, which selects questions with high informativeness and representativeness. By incorporating the structural information of knowledge concepts, KG-EIR recommends exercises to students, thereby improving their overall cognitive level during the recommendation process.
	\item We design four innovative components and a novel cognitive diagnosis model, NACD, including the informativeness component, exercise representation component, knowledge importance component, and exercise representativeness component. The informativeness component estimates the informativeness of each exercise and generates the candidate question set. This question set serves as input to the exercise representativeness component, which selects questions with high knowledge coverage based on the knowledge importance component. The knowledge importance component incorporates a multidimensional KG and a knowledge points extraction algorithm with five skill features to generate skill importance weights. Finally, the cognitive diagnosis model predicts student performance and estimates their current knowledge state based on exercise and skill relations.
	\item We evaluate the KG-EIR framework on two public educational datasets, including Assistment 2009-2010 and Eedi 2020, using the AUC (Informative Metric) and Knowledge Coverage Rate (KCR). The experimental results demonstrate that KG-EIR outperforms other methods such as RAND and EM.
\end{enumerate}

The rest of the paper is structured as follows. Section 2 provides an overview of related works on cognitive diagnosis models, relation modeling, and exercise recommendation. Section 3 presents important terminologies, defines the goal and problem statement of this study. Section 4 describes the methods proposed in this paper. Section 5 presents the experimental evaluation of our framework using two different metrics. Section 6 concludes the paper and discusses future research directions.

\section{Related Work}\label{sec2}

\subsection{Cognitive Diagnosis Model}
Cognitive diagnosis plays a crucial role in various real-world scenarios, including games \cite{chen2016predicting}, medical diagnosis \cite{xu2017predicting}, and especially education \cite{liu2018fuzzy}. The primary objective of cognitive diagnosis is to uncover the latent trait characteristics of learners based on their testing records. These discovered characteristic features have applications in tasks such as resource recommendation \cite{chen2018recommendation} and performance prediction \cite{wang2020neural}. Early approaches to cognitive diagnosis mainly relied on psychological evaluation \cite{liu2021towards}. The two most traditional cognitive diagnosis models, namely the Item Response Theory (IRT) \cite{lord1952theory} and the Deterministic Input, Noisy And Gate (DINA) model \cite{de2009dina}, model the response generated by a learner answering an item as the interaction between the learner's trait features and the item. Ackerman et al. \cite{ackerman2014multidimensional} extended the characteristic features into a multidimensional space by proposing the Multidimensional Item Response Theory (MIRT). In recent years, deep learning has been incorporated into cognitive diagnostics by several researchers \cite{tsutsumi2021deep,wu2020variational}. Wang et al. \cite{wang2020neural} introduced NeuralCD, which utilizes neural networks to autonomously learn the interaction function. However, these cognitive diagnosis models overlook the deep relations between exercises, skills, and students when estimating students' knowledge state.

\subsection{Relation Modeling}
Based on psychological research, the relationship between exercises and skills has been extensively explored in numerous studies that measure students' knowledge levels (e.g., \cite{song2020sepn}, \cite{yang2021gikt}). Many researchers employ Q-matrices to model the relationship between exercises and skills, where exercises related to the same knowledge concept are considered connected in the Q-matrix. Additionally, researchers investigate the relationship between two exercises or skills based on exercise embeddings (e.g., \cite{nagatani2019augmenting,huo2020knowledge}). Semantic similarity scores of exercises are computed using prior interactions to model the significance of these interactions. However, these relation modeling methods do not consider the heterogeneous interactions between students, exercises, and skills. Therefore, this paper incorporates knowledge graphs (KGs) and Graph Convolutional Networks (GCNs) to establish exercise and skill relations and delve into exercise features such as informativeness and representativeness.

\subsection{Exercise Recommendation}
Traditional recommendation systems employ collaborative filtering, which can be categorized into nearest-neighbor collaborative filtering and model-based collaborative filtering. Nearest-neighbor collaborative filtering includes user-based collaborative filtering \cite{chuan2006recommendation} and item-based collaborative filtering \cite{kim2006new}. Model-based collaborative filtering involves mining hidden or explicit features to mitigate data sparsity and achieve good scalability \cite{koren2009matrix}. When applying traditional recommendation methods to exercise recommendation in the educational field, students are treated as users and exercises as items. Thus, nearest-neighbor collaborative filtering can be further classified as exercise-based and student-based. Taking the impact of knowledge graphs into account, recent research has proposed knowledge graph-based recommendation methods for exercise recommendations (e.g., \cite{shi2020learning} and \cite{zhu2018multi}).

Recent exercise recommendation methods that leverage knowledge graphs help mitigate misunderstandings in learning content descriptions. Inspired by this idea, Wan et al. \cite{wan2016learner} introduced a learner-oriented exercise recommendation method based on knowledge concepts, represented as nodes, and the relationships between them as edges in knowledge graphs. Ouf et al. \cite{ouf2017proposed} developed exercise recommendation methods by incorporating knowledge graphs with semantic web ontologies to merge personalized concepts. To organize learning resources in a sequential manner, Shmelev et al. \cite{shmelev2015approach} proposed a method that integrates evolutionary methods and knowledge graph technology. Chu et al. \cite{chu2011ontology} created an e-learning system based on a conceptual map that can generate learning paths using the connections in the concept map. Recognizing the need for diverse learning paths in different settings, Zhu et al. \cite{zhu2018multi} presented a method for recommending learning paths using pre-built learning scenarios. They developed an approach that requires the definition of starting and ending nodes to construct learning paths.

\section{Preliminaries}\label{sec3}
This section is divided into three parts. The first part presents the problem addressed in this paper. The second part provides definitions for several terminologies used throughout the paper, including exercise informativeness, exercise representativeness, and heterogeneous interactions. The third part outlines the goals of the paper.

\begin{table}[!htbp] 
	\caption{Important mathematical notations.\label{tab1}}
	\centering
	\begin{tabular}{c|c}
		\hline
		\textbf{Notations}	& \textbf{Descriptions }\\ 
		\hline
		KC & The knowledge concepts \\
		KG & The knowledge graph \\
		P & The learning paths in the knowledge graph\\
		$N_e$ & The total number of exercises\\
		K & The total number of skills\\
		$Q_U$ & The untested question set\\ 
		$Q_C$ & The candidate question set\\
		$Q_T$ & The tested question set \\
		$Level(KC)$ & The level of KC \\
		$P_n$  & Response matrix\\ 
		$\hat{E}$  & Dissimilarity matrix \\ 
		\bottomrule
	\end{tabular}
\end{table}

\subsection{Problem Definition}
Exercise recommendation aims to suggest exercises that enhance students' knowledge proficiency. The problem at hand is how to recommend appropriate exercises that meet the specific requirements of each student. In this paper, two measurements are defined to evaluate exercise quality: exercise representativeness and exercise informativeness. Thus, the specific problem addressed in this paper is how to recommend exercises with high representativeness and informativeness from a large pool of questions. To solve this problem, we propose the Knowledge-Graph-Exercise Informativeness and Representativeness (KG-EIR) framework, which comprises four components and a cognitive diagnosis model. Specifically, the informativeness component selects exercises with high informativeness from the untested question set ($Q_U$) to the candidate question set ($Q_C$). The exercise representation component and the knowledge importance component are designed to generate skill and exercise embeddings, as well as skill importance weights. The outputs of the exercise representation component and the knowledge importance component serve as input to the exercise representativeness component, which selects questions with high representativeness from $Q_C$ to the tested question set ($Q_T$). Finally, the Neural Attentive Cognitive Diagnosis (NACD) model predicts students' performance on $Q_T$ and diagnoses their current knowledge state.

\subsection{Terminologies}

\paragraph{Definition 1: Informativeness} In general, a valid question is expected to reduce the level of uncertainty in an examinee's knowledge state. Thus, the informativeness of an exercise can be defined as the amount of information that the underlying cognitive diagnosis model (M) can acquire from the question to update the estimate of knowledge states. Selecting the most informative questions is a means of achieving the informativeness goal. After the student completes the test, the performance of the student with M on the entire tested question set is predicted, and the performance is evaluated using a metric such as the Area Under the Curve (AUC), denoted as Inf(S).

\paragraph{Definition 2: Representativeness} If a set of questions achieves a certain knowledge coverage rate, it is considered representative. The knowledge coverage rate is used as a measure of representativeness. Selecting a group of questions with the highest coverage of knowledge concepts is essential to achieve the representativeness objective. The coverage, Cov(S), can be computed as the percentage of knowledge concepts covered by the tested question set, $Q_T$, after the test.

\begin{figure}[!htbp]
    \centering
    \includegraphics[width = 0.5\textwidth,height = 7cm]{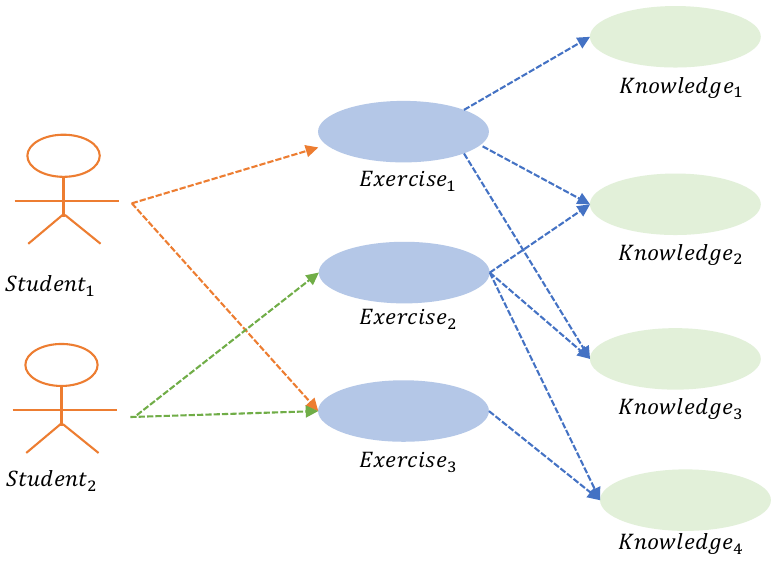}
    \caption{Toy example illustrating the heterogeneous interaction between students, exercises, and skills.}
    \label{hi}
\end{figure}

\paragraph{Definition 3: Heterogeneous Interaction} When answering exercises, there exist various interactions among students, exercises, and skills. Heterogeneous interaction, denoted as HI = (V; E), consists of an object set, V, and a link set, E. The object types in V include students, exercises, and skills. E is a collection of relational types in the form E = ($r_A$, $r_C$), where $r_A$ represents the relation it answers and $r_C$ represents the relation it contains. Figure \ref{hi} provides a toy example illustrating this definition.

\subsection{Goals}
The goal of this paper is to recommend exercises with high representativeness and informativeness to improve student performance in subsequent interactions. Informativeness is measured using the Area Under the Curve (AUC), while representativeness is measured by the knowledge coverage rate when predicting the corresponding exercises.

\section{Proposed Method}\label{sec4}

\begin{figure}[!htbp]
	\centering
	\includegraphics[width=\textwidth]{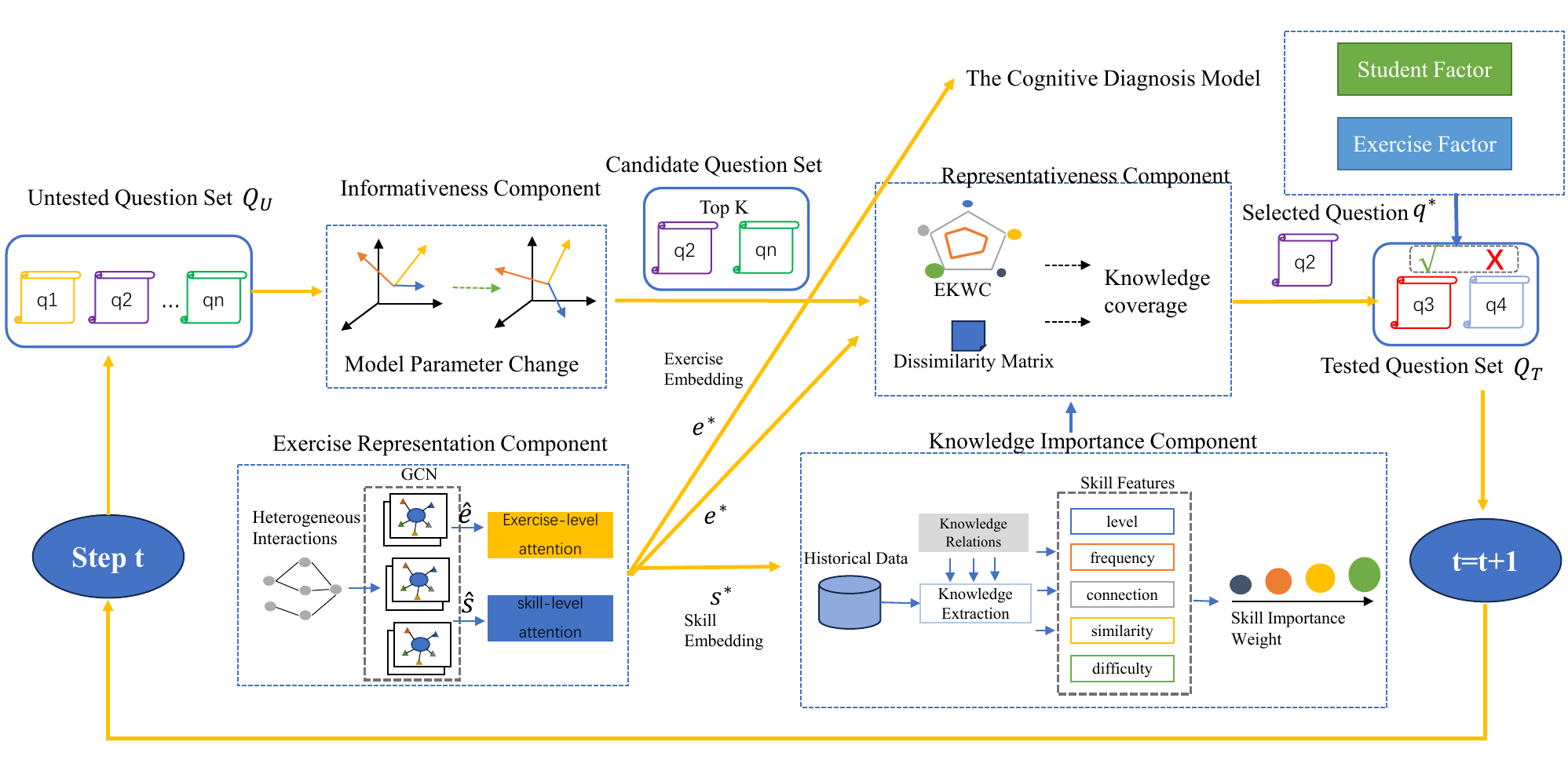}
	\caption{The overall framework of the KG-EIR selection strategy. The framework consists of four components: the informativeness component, the exercise representation component, the exercise representativeness component, and the knowledge importance component. The informativeness component selects high-informativeness questions to form the candidate question set $Q_C$. The exercise representation component extracts exercise and skill embeddings through a graph neural network. The exercise representativeness component selects a question with high representativeness to maximize the marginal gain. The knowledge importance component assesses the relevance of knowledge concepts. The NACD model predicts student performance based on the exercise embedding.}
	\label{overall}
\end{figure}

In this section, we present our proposed framework called Knowledge Graph-Exercise Informativeness and Representativeness (KG-EIR). The framework aims to model exercise features and skill features to generate questions based on their informativeness and representativeness. KG-EIR consists of four components: the informativeness component, the exercise representation component, the exercise representativeness component, and the knowledge importance component. The overall structure and components of KG-EIR are depicted in Figure~\ref{overall}.

The KG-EIR framework operates as follows: Given a user $e_i \in E$, the framework models exercise features and skill features to generate questions that are both informative and representative. At each step $t$, KG-EIR selects one question from the untested question set $Q_U$ to be added to the tested question set $Q_T$. The framework can be divided into four components:

\textbf{Informativeness Component}: This component is responsible for selecting a candidate question set $Q_C$ from $Q_U$ based on informativeness. The selection is performed using a score function called Model Parameter Change (MPC). MPC estimates the user's knowledge states by observing their answers and quantifies the extent to which a question alters the diagnosis. The top-K highly informative questions are selected to form $Q_C$.

\textbf{Exercise Representation Component}: In this component, the framework extracts information on the heterogeneous interactions between users, exercises, and skills using a graph neural network. The exercise representation component generates the exercise embedding $e^*$ and skill embedding $s^*$. These embeddings serve as inputs for the subsequent components.

\textbf{Exercise Representativeness Component}: The representativeness component selects a question with high representativeness from $Q_C$ to maximize the marginal gain. The selection process takes into account the exercise embedding $e^*$ obtained from the exercise representation component.

\textbf{Knowledge Importance Component}: To enhance the selection process, the knowledge importance component assesses the relevance of knowledge concepts. It explores the relevance of knowledge concepts using the knowledge points extraction algorithm and incorporates five skill features to generate the skill importance weight.

Finally, the NACD model predicts student performance and estimates their state based on the exercise embedding $e^*$.

\subsection{Informativeness Component}
The informativeness component is the first step of the KG-EIR framework, where we select a candidate question set $Q_C$ consisting of top-K high-informativeness questions. To measure the informativeness of a question, we propose a score function called Model Parameter Change (MPC).

MPC leverages the information contained in a question by estimating the user's knowledge states after answering the question. The parameter change in the abstract Cognitive Diagnosis Model (CDM), denoted by $\theta$ in M, represents the knowledge states in KG-EIR. The amount of change in the CDM parameters reflects the amount of information gathered from the question. If the $\theta$ values change significantly, the question is considered more informative; otherwise, it provides little information.

The MPC function calculates the probability of correctly answering a question, which can be predicted by the cognitive diagnosis model M. Let $\bigtriangleup M = |\theta(R \cup {r_{ij}}) - \theta(R)|$ represent the parameter change in our model when adding the record $r_{ij} = <e_i, q_j, a_{ij}>$. Here, $\theta(R_i)$ represents the parameters obtained from the current interaction $R_i$ of student $e_i$, and $\theta(R_i \cup {r_{i,j}})$ represents the parameters after adding the interaction. For each question $q_j$, the MPC function is defined as follows:

\begin{eqnarray}
    EMC(q_j) &=& E_{a_{ij}~p} \bigtriangleup M(<e_i,q_j,a_{i,j}>)\\
    p&=&M(e_i,q_j|\theta(R_i))
\end{eqnarray}

The $\bigtriangleup M(r_{ij})$ is approximated by the gradient caused by $r_{ij}$. This approach is particularly effective for models trained using gradient-based methods, such as neural models.

Based on the MPC score function, we select questions from the untested question set to form the candidate question set $Q_C$. We calculate the MPC for each question and select the top-K questions with the highest informativeness.

\subsection{Exercise Representation Component}
The exercise representation component is the second step of the KG-EIR framework, where we extract exercise embedding ($e^*$) and skill embedding ($s^*$) by considering the heterogeneous interactions between students, exercises, and skills.

We employ the Graph Convolutional Network (GCN) model to generate embedding representations of exercises and skills, capturing their static relationships. Before applying the GCN model, we define the neighbors of exercises and skills based on three meta-relationships: exercise-student-exercise (eSe), exercise-skill-exercise (eKe), and skill-exercise-skill (kEk). In the eSe and eKe relationships, the exercise neighbors are exercises answered by the same student or covering the same skill. In the kEk relationship, the skill neighbors are skills contained in the same exercise. To propagate information in the GCN, we use two matrices: the exercise relation matrix ($R^E$) and the skill relation matrix ($R^S$), which capture the high-order information. Then, we apply the GCN model to generate the hidden embedding representations of exercises ($\hat{e}$) and skills ($\hat{s}$).

Each convolutional layer in the GCN model updates the nodes based on their own state and the state of their nearest neighbors. Let $node_i$ denote the state of an exercise or skill, and $Node(i)$ denote a group of nodes representing the neighbors of $node_i$. The exercise at the $\emph{l}$-th layer can be computed as follows:
\begin{equation}
    node^{\imath}_i = RELU(\sum_{j\in {i}\cup Node(i)}^{} w^{\imath}_{i}node^{\imath-1}_j + b^{\imath}_i)
\end{equation}
where $w^\imath$ and $b^\imath$ represent the weight matrix and bias of the GCN layer, respectively, and $\text{RELU()}$ denotes the activation function used in the GCN model.

The hidden embedding representations obtained from the GCN model capture the static relationships between exercises and skills. However, they do not consider the similarity among exercises and skills when generating their embeddings. To incorporate the deep semantics of exercises and skills, we use exercise-level attention and skill-level attention mechanisms. These attention mechanisms learn the semantic relationships between students and exercises, generating the final embedding representations $e^*$ and $s^*$. The formulation for $e^*$ is as follows:
\begin{eqnarray}
   \alpha_E  &=&  softmax(\frac{(\hat{e}W^Q)(\hat{e}W^K)}{\sqrt{d_K}})\\
   \beta_E  &=&  \delta_a  \alpha_E +(1-\delta_a  )R^E \quad  e^* = \beta_E  \hat{e}W^v
\end{eqnarray}
Here, $R^E$ is the exercise relation matrix, $\sqrt{d_K}$ is the scaling factor, and $W^K$, $W^Q$, and $W^V$ are projection matrices.

The process for obtaining $s^*$ is similar to that of $e^*$. The difference lies in using the hidden embedding representation of skills as input to the attention mechanism, and the skill relation matrix is used instead of the exercise relation matrix when calculating $\beta_E$.

\subsection{Knowledge Importance Component}

After the procedure of selecting questions from the untested question set to the candidate question set, the Knowledge Importance component aims to compute the knowledge importance weight $W_K$ as input for the next selection procedure: the representativeness component. Previous studies \cite{giunchiglia2009faceted} have shown that organizing educational resources into different classes helps students understand learning profiles and enables them to logically organize and recall knowledge. Therefore, in our knowledge graph, we separate knowledge concepts into different classes to learn the weight of knowledge concepts (KCs).

We categorize learning objects into three classes:
\begin{itemize}
	\item Subject Knowledge: This class contains KCs at the subject level, such as "math," "physics," and "biology," supporting basic knowledge areas like "Ratio," "Geometry," and "Standard Form."
	\item Basic Knowledge: The core of the framework, this class includes specific knowledge fields such as "Proportion" and "Negative Numbers" that are essential for solving specific tasks.
	\item Task: This class encompasses practical educational problems like "Factorising into a Single Bracket" and "Expanding Single Brackets." The task level is the bottom level of our knowledge graph framework.
\end{itemize}

\begin{figure}[!htbp]
    \centering
    \includegraphics[width=\textwidth]{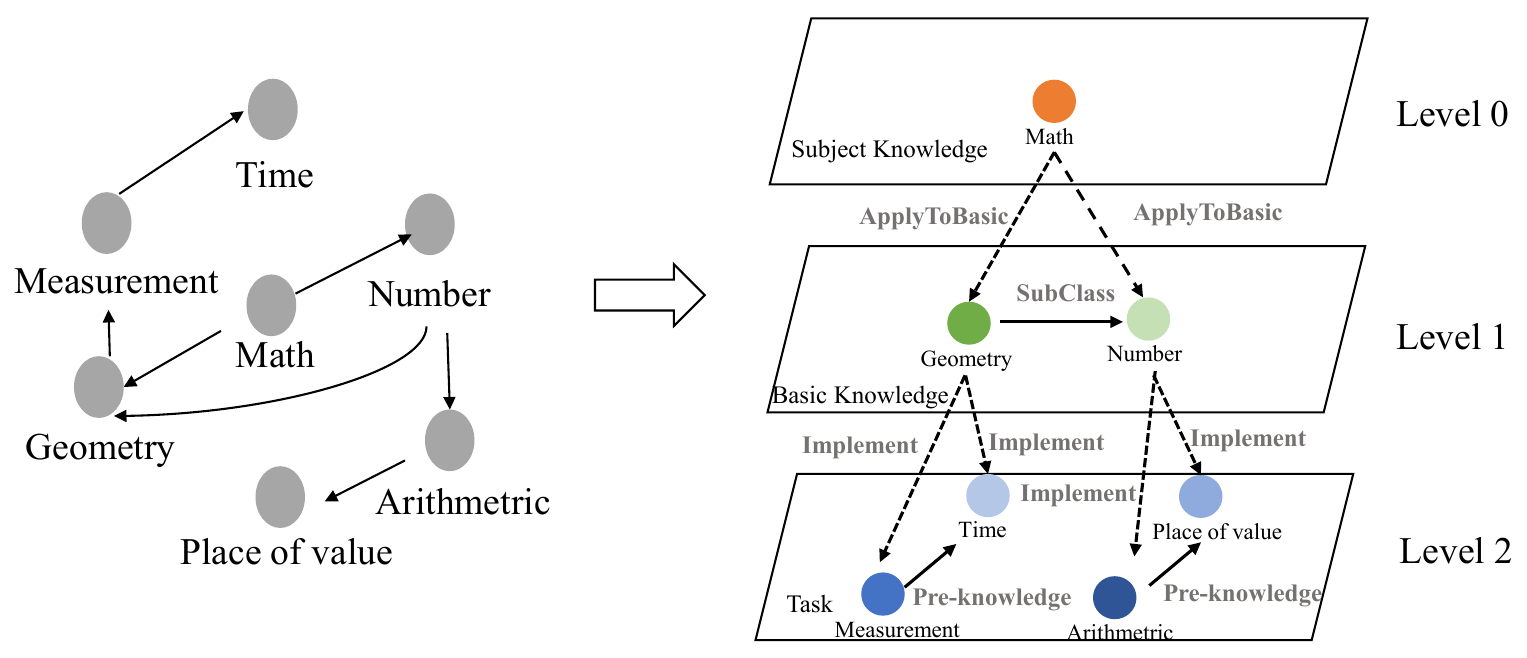}
    \caption{(Left) One-dimensional KG framework used in previous studies. (Right) Multidimensional KG framework employed in our paper. Dotted lines represent links between classes, solid lines denote interactions within classes, and nodes of various colors represent learning items in different classes.}
    \label{kl}
\end{figure}

Figure~\ref{kl} presents a visual representation of the multidimensional KG framework we employ in our paper. Each class in this framework consists of a hierarchy and associated learning object instances. The learning objects represent meta-learning resources that are incorporated into the hierarchy and connected by semantic relationships, while the hierarchy reflects the knowledge structure of the current class. We establish various relationships, dividing them into intraclass relationships and interclass interactions, to illustrate the semantic connections between learning objects. Intraclass relationships link learning objects within a class, while interclass relationships provide links between educational resources from different classes (see Table~\ref{q_statistic}). Our knowledge graph expands the accessibility of learning objects across classes and strengthens connections between cross-class learning objects. This graph of knowledge showcases how learned information can be practically applied, deepening e-learners' understanding of the studied information and helping them comprehend how theoretical knowledge can be used in practical scenarios.

\begin{table}[!htbp]
	\centering
	\caption{Designed relationships in the knowledge graph.\label{q_statistic}}
	\begin{tabularx}{\textwidth}{XXX}
		\toprule
		\textbf{Knowledge Relationship} & \textbf{Type} & \textbf{Description} \\
		\midrule
		Subclass & Intra-class & Indicates that the current LO possesses a subclass. \\
		Implement & Inter-class & Indicates that the current LO can implement subsequent LO. \\
		Pre-knowledge & Intra-class & Indicates that prior knowledge exists that should be learned before the current LO (basic knowledge). \\
		ApplyToBasic & Inter-class & Indicates that the current LO can be applied in the target LO. \\
		\bottomrule
	\end{tabularx}%
\end{table}%

\subsubsection{Knowledge Points Path Extraction Algorithm}

To determine the relevance of KCs, we explore all possible learning paths through the target learning object and the learning need of the e-learner. We have designed the Knowledge Points Path Extraction Algorithm, which is based on the multidimensional knowledge graph, to accomplish this task. The algorithm consists of two phases.

The first phase involves calculating the relationship constraints $\phi$ based on the learning need. The \emph{getRelation()} function is used to determine the relationship constraints $\phi = (\alpha, \beta, \gamma...)$ corresponding to the learners' needs.

In the second phase of the algorithm, a learning path is constructed using the relationship restrictions. Starting with the target learning item, the algorithm generates the learning path by searching for the next learning object associated with a relationship that satisfies the constraints. The associated learning object serves as a continuation of the search. The initial learning object of the current learning route is the chosen target learning object. If the current learning item has no related learning objects, the path will consist of only one KC. The algorithm performs a greedy search starting from the target learning object.

For a detailed description of the algorithm, please refer to Algorithm~\ref{ag1}.
\begin{algorithm}[!htbp]
\DontPrintSemicolon
  \SetAlgoLined
  \KwIn {Students' historical response dataset: $D = {s_1, s_2,...s_N}$, $s_i=(e_i, s_i, t_i)$;The knowledge level graph G;}
  \KwOut {The all possible learning path: P.}
  
  \While{findAllPaths($KC_u$)}{
  R  = getRelations($KC_u$) (find the relations connected with $KC_u$) \; 
  \uIf {$\left \{ r| r\in R, r\in \oslash  \right \}$} {
    P.addPath(p);{Add the new path into the path set}
} \lElse {
   \While{all r $\in$ R}{
   $list(KC_u)$ = getConnectObject($KC_u$) (obtain the connected objects of $KC_u$ with r)\;
   p.addElement($list(KC_u)$) (put the element in the learning path)\;
   Recursively apply findAllPaths($KC_u$);
  } 
} }
  \caption{Knowledge Points Extraction Path Algorithm.}
  \label{ag1}
\end{algorithm}

\subsubsection{Knowledge Importance Weight Extraction Algorithm}

The Knowledge Importance Weight Extraction Algorithm aims to extract the weights of KCs based on five skill features. In previous work on quantifying algorithms \cite{mester2016rankings}, the feature set $F$ of KCs was proposed to select important KCs, including the level ($f_1$), frequency ($f_2$), connection ($f_3$), similarity ($f_4$), and difficulty ($f_5$) of the corresponding KCs.

\begin{itemize}
    \item The level feature ($f_1$) is designed to extract the level of a KC. By applying the Knowledge Points Path Extraction Algorithm (KPE), which transforms the one-dimensional knowledge graph into a multi-dimensional one, the levels of KCs in all related learning paths can be extracted. For example, if the output of the KPE is "A-B-C," where the levels of A, B, and C are 0, 1, and 2, respectively, the different knowledge levels of KCs can be extracted based on different learning paths. The total level of a KC can be defined as follows:
    
    \begin{equation}
        f_1(KC) = \frac{\sum_{i=0}^{N_{C}} Level(KC)_{i}}{N_{C}}
    \end{equation}
    where $N_C$ is the number of learning paths that contain the KC.
    
    \item The frequency feature ($f_2$) focuses on extracting the frequency of a KC in all learning paths. A greedy algorithm is used to search for the frequency of the KC across all learning paths. The frequency of the KC can be defined as follows:
    \begin{equation}
        f_2(KC) = \frac{N_C}{N}
    \end{equation}
    where $N$ indicates the total number of learning paths.
    
    \item The connection feature ($f_3$) considers the connections between KCs. When KCs occur in the same learning path, they are considered connected. For example, in a learning path "A-B-C," skill A is connected to skills B and C. The connectivity can be calculated as follows:
    \begin{equation}
     f_3(KC) = \frac{|ConnectSet(KC)|}{K}
    \end{equation}
    where \emph{ConnectSet} represents the list of connected KCs, and $K$ is the total number of KCs.
    
    \item The similarity feature ($f_4$) is used to explore the similarity between each skill. It utilizes the skill representations ($s^*$) and calculates the dot product to measure the similarity between skills: 
    \begin{equation}
    f_4(KC) = \frac{s^*_i\cdot  s^*_j}{|s^*_i||s^*_j|}
    \end{equation}

    \item The difficulty feature ($f_5$) leverages students' interactions to indicate the difficulty of skills. It models the cognitive difficulty of skills based on students' behavior when they attempt exercises containing the same skill at different timestamps. The cognitive difficulty of a skill set for each student ($S_i$) at timestamp $t$ is represented by $\pi_{S_i, KC, t}$:
     
\begin{align}
\begin{split}
\pi  _{S_i,KC,t}= \left \{
\begin{array}{ll} 
     \left [ \frac{|\{A_s==0\}|_{0:t}}{|Q|_{0:t}}*4  \right ] &  if|N_v|_{0:t}\ge 5  \\          
     5                                 & otherwise
\end{array}
\right.
\end{split}
\end{align}
Here, ${A_s = 0}$ represents the set of questions where the student answered incorrectly for the questions containing the KC. The cognitive difficulty of the KC is divided into five levels if a learner has performed fewer than five attempts to answer the question. The average cognitive difficulty for different learners on the KC is defined as the difficulty feature of the KC:

\begin{equation}
    f_5(KC) = \frac{\sum_{i=0}^{N_S}\sum_{j=0}^{N_T}\pi_{S_i,KC,t_j}}{N_S\times N_T}
\end{equation}
where $N_S$ is the number of students and $N_T$ is the time consumption.
\end{itemize}

In this paper, we consider the novelty and popularity of KCs. To satisfy different learning preferences, we apply a weighted method ($W$) to combine the five skill features using the following equation (Equation~\ref{w}). Each weight ($w$) in our learning preference options corresponds to a certain feature ($f_i$). The weight distribution details are as follows:
\begin{itemize}
    \item Novelty: When considering the novelty of exercises, we set $w_1 = 0.5$, $w_2 = 0$, $w_3 = 0$, $w_4 = 0$, and $w_5 = 0.5$. We consider the level and difficulty of skills to represent the inherent novelty of exercises.
    \item Popularity: When considering the popularity of exercises, we set $w_1 = 0$, $w_2 = 0.6$, $w_3 = 0.1$, $w_4 = 0.3$, and $w_5 = 0$. We consider the frequency, connection, and similarity of skills to model the popularity of skills.
\end{itemize}

\begin{equation}
    W = \sum_{i=0}^{5} W_i \times f_i(KC)
    \label{w}
\end{equation}

The weighted method calculates the weight of novelty ($W_{nov}$) and popularity ($W_{pop}$) for each KC.

Finally, by combining novelty and popularity, the skill importance weight ($W_K$) is obtained. The formulation is as follows:
\begin{eqnarray}
    W_{skill} = W_{nov} + W_{pop} \\
    W_{K} = Tanh(W_{skill})
\end{eqnarray}
where $\tanh = \frac{e^z - e^{-z}}{e^z + e^{-z}}$.

The Knowledge Importance Weight Extraction Algorithm allows us to determine the weights of KCs based on their skill features, incorporating novelty and popularity considerations. These weights play a crucial role in assessing the importance of KCs within the learning context.

\subsection{Exercise Representativeness Component}

After collecting a candidate set $Q_C$ of highly informative questions and obtaining skill importance weights and exercise embeddings, this section focuses on designing the exercise representativeness component. The goal is to select questions from $Q_C$ into the tested question set $Q_T$ that exhibit high representativeness. To assess the informativeness of exercises, a novel scoring function is proposed to evaluate the knowledge coverage of $Q_T$. An approach is then devised to gradually add more questions to $Q_T$ until it achieves the highest coverage score.

The knowledge coverage of the tested question set $Q_T$ can be estimated by checking whether the corresponding KCs exist in $Q_C$. Therefore, a straightforward knowledge coverage function, denoted as SKC, is designed as follows:
\begin{eqnarray}
    Cov(KC, Q_c) &=& 1  \quad when \quad \exists KC \in Q_C\\
    SKC(Q_c) &=& \frac{\sum_{i=0 }^{K}Cov(KC,Q_c)}{|K|}
\end{eqnarray}
Here, $Cov(KC, Q_C) = 1$ indicates that the KC is involved in $Q_C$. However, SKC has two obvious flaws. Firstly, it considers all KCs equally and fails to distinguish the importance of each KC. Secondly, the value of $Cov$ is binary and does not reflect the number of exercises. For example, if the math quiz focuses on "Fractions" rather than "Real Numbers," it is more appropriate to select more fractions-related problems rather than simply covering both topics equally. Choosing nine questions about "Real Numbers" and one question about "Fractions" should be equivalent to choosing five questions from each.

To address these flaws, the exercise weight knowledge coverage function (EWKC) is proposed to calculate the knowledge coverage of the tested question set $Q_T$. Specifically, the EWKC function combines the number of exercises to generate the knowledge coverage of $Q_C$. Moreover, to account for the importance of exercises and KCs, skill importance weights obtained from the Knowledge Importance Component are incorporated. The EWKC function is defined as follows:
\begin{eqnarray}
    cnt(KC, Q_T) &=& \sum_{q \in Q_C}^{} 1 [(q,KC) \in Q_T] \\
    ECov(KC, Q_T) &=& \frac{cnt(KC, Q_T))}{1+e^{(-cnt(KC, Q_T))}} \\
    EWKC(Q_T) &=& \frac{\sum_{k\in K}^{}W_kECov(k, Q_T)}{\sum_{k\in K}W_k} \label{ewkc} 
\end{eqnarray}
Here, $W_K$ represents the exercise weight for the concept $k$, which is discussed in the Knowledge Importance Component. The ECov function counts the occurrence of a KC in $Q_T$ and applies a sigmoid function to ensure the coverage value lies within the range of 0 and 1. Finally, the EWKC function calculates the weighted average knowledge coverage over all KCs, with weights determined by the importance of the corresponding skills.

However, the EWKC function only considers the impact of skills and ignores the influence of exercises. To better define the representativeness of exercises, the Response Matrix ($P_n$) and Dissimilarity Matrix ($\hat{E}$) are introduced.

\subsubsection{Response Matrix}

The Response Matrix $P_n$ of size $|S| \times N_e$ is designed, where each element is defined as follows:
\begin{align}
\label{pn}
\begin{split}
P_n (i,j)= \left \{
\begin{array}{ll} 
       a_j^i   &  if j \le  N_e\\          
     0                                 & otherwise
\end{array}
\right.
\end{split}
\end{align}
The matrix $P_n$ stores the probability of students answering the next exercises correctly. $|S|$ represents the number of students, $|C|$ represents the number of exercises done by each student, and $N_e$ represents the total number of exercises. If an exercise was not done by a student, the corresponding columns are filled with zeros. These columns correspond to the $N_e - |C|$ hypothetical exercises that students cannot answer correctly and will be replaced by other exercises in the future.

\subsubsection{Dissimilarity Matrix}

To consider exercise representativeness, the dissimilarity between exercises, denoted as $\hat{E}$, is defined as follows:
\begin{equation}
    \hat{E}_{ij} = 1 - \frac{\hat{e^*_i}\cdot  \hat{e^*_j}}{|\hat{e^*_i}||\hat{e^*_j}|}
\end{equation}
Here, $\hat{e^*}$ represents the exercise representation based on the exercise representation component.

The final knowledge coverage combines skill features and exercise features to measure the representativeness of exercises. It is defined as follows:
\begin{equation}
    R_{ij} = \alpha_1 \sum_{KC \in Q_T} EWKC(KC) + \alpha_2 P_{n}(i,j) + \alpha_3 \hat{E}_{ij}
\end{equation}
where $\alpha_1$, $\alpha_2$, and $\alpha_3$ are hyperparameters in the model. The representativeness of exercises is evaluated based on the weighted sum of the knowledge coverage of KCs in $Q_T$, the probabilities stored in the response matrix $P_n$, and the dissimilarity matrix $\hat{E}$. 

The exercise representativeness component aims to select questions from the candidate set $Q_C$ into the tested question set $Q_T$ with high representativeness. The knowledge coverage of $Q_T$ is evaluated using the EWKC function, which incorporates skill importance weights and exercise numbers. The response matrix $P_n$ and dissimilarity matrix $\hat{E}$ are introduced to consider exercise features and representativeness. The final knowledge coverage is calculated by combining skill and exercise features. The hyperparameters $\alpha_1$, $\alpha_2$, and $\alpha_3$ control the relative importance of these features in measuring exercise representativeness.

\subsection{Cognitive Diagnosis Model}

This section introduces a novel cognitive diagnosis model, NACD, within the KG-EIR framework. The NACD model aims to estimate the knowledge state of students and make predictions on the tested question set. To achieve accurate diagnosis, the NAKT model incorporates student factor modeling and exercise factor modeling. The student factor modeling focuses on capturing students' behavior during exercise training, specifically their slipping behavior and guessing behavior. Additionally, the exercise factor is modeled based on the output $e^*$ generated by the exercise representation component. The exercise factor aims to explore the relationship between exercises and skills and utilizes the exercise-skill relation matrix as input for a relative-distance attention mechanism to generate the exercise factor representation.

\subsubsection{Exercise Factor}

To model the relationship between exercises and skills, a binary Q-matrix $Q$ is constructed to map exercises to skills. If exercise $e_i$ contains a knowledge point $k$, the corresponding entry $Q_{i,k}$ is set to the skill importance weight $W_K$. Otherwise, $Q_{i,k}$ is set to zero. Based on the Q-matrix, the knowledge point vector of exercise $e$ can be obtained as follows:

\begin{equation}
    K^V = x^e \times Q^T
\end{equation}

Here, $Q \in \mathbb{R}^{N_e \times K}$, and $x^e \in {0,1}^{N_e \times 1}$ represents the one-hot representation of exercises. The exercise embedding, composed of the corresponding $K^V$ values, is then used as input for the relative distance mechanism. The relative distance between input sequences, represented by $x_i$ and $x_j = (x_1, x_2, \ldots, x_{n-1})$, is captured using edge vectors $a_{i,j}^V$ and $a_{i,j}^K$. To prevent unbounded values, the edge vectors are clipped using the function $clip(x, k) = \max(-k, \min(k,x))$, where $k$ represents the maximum absolute value. The associated relative position representations for $W^K$ and $W^V$ are defined as $W^K = (W^k_{-k}, \ldots, W^K_{k})$ and $W^V = (W^V_{-k}, \ldots, W^V_k)$. Finally, the relative position attention mechanism outputs the exercise factor representation $F^E$. The following equations describe the process:

\begin{eqnarray}
    a^k_{i,j}  &=&  W^k_{clip}(j-i,k) \\
    a^V_{i,j}  &=&  W^V_{clip}(j-i,k)  
\end{eqnarray}

The edge vectors are then utilized as input for the attention mechanism. The attention weights $a_{i,j}$ are calculated based on the relative distances $e_{i,j}$, which are computed as:
\begin{eqnarray}
    a_{i,j} &=& \frac{exp(e_{i,j})}{\sum_{i=1}^{n}exp(e_{i,k}) } \\
    e_{i,j} &=& \frac{x_iW^Q(x_jW^k)^T + x_iW^Q(a_{i,j}^K)^T}{\sqrt{d_{F^E}}}\\
     F^E_i &=& \sum_{j=1}^{n} a_{ij}(x_jW^V)
\end{eqnarray}
In the above equations, $W^Q$, $W^K$, and $W^V$ represent the query, key, and value matrices, respectively, and $d_{F^E}$ denotes the dimension of $F^E$.

\subsubsection{Student Factor}

The student factor, denoted as $F^S$, models the representation of students based on their knowledge proficiency vectors in diagnosing their states. The formulation for the student representation is as follows:

\begin{equation}
    H^S = sigmoid(x^s \times A)
\end{equation}
Here, $x^S \in {0,1}^{S \times 1}$ represents the one-hot encoding of students, and $A$ is a trainable matrix within the framework.

Next, we introduce two factors related to student behavior: the slipping factor and the guessing factor. The slipping factor captures situations where a student attempts to complete an exercise but provides an incorrect answer due to careless mistakes. The guessing factor represents instances when a student may guess an answer because they have not fully mastered the corresponding skills. The formulations for the slipping factor and the guessing factor are as follows:

\begin{eqnarray}
    H^{Slipping} = sigmoid(x^e \times B)\\
    H^{Guessing} = sigmoid(x^e \times C)
\end{eqnarray}
Here, $B$ and $C$ are trainable matrices.

To generate the student factor representation $F^S$, we incorporate $H^S$, $H^{Slipping}$, and $H^{Guessing}$ as inputs to a two-layer linear network. The input for the linear network is defined as follows:

\begin{equation}
     X = Q \times (H^{S} - H^{Slipping}) \times H^{Guessing}
\end{equation}

Subsequently, $X$ serves as the input for the linear network:
\begin{eqnarray}
    f_1 &=& \sigma (W_1\times X + b_1) \\
    F^S &=& \sigma (W_2\times f_1 + b_2)
\end{eqnarray}
Here, $\sigma()$ represents the sigmoid activation function.

\subsubsection{Student Performance Prediction.} 

To predict student performance on a tested question set, $Q_T$, we combine the student factor $F^S$ and the exercise factor $F^E$. The formulation for this prediction is as follows:

\begin{equation}
    p = sigmoid(W_s\times F^S+W_e\times F^E +b_p)
\end{equation}
Here, $W_s$ and $W_e$ are weighted matrices, $b_p$ is the bias vector, and $p$ represents the likelihood that the student will answer the subsequent interaction exercise, denoted as $e_{N_e+1}$, correctly.

\section{Experiments}\label{sec5}

In this section, we conduct experiments using two public educational datasets: the Assistment dataset and the Eedi dataset to investigate the performance of our selection strategy: KG-EIR. The experiments are organized into five aspects. First, we compare the novel cognitive diagnosis model, NACD, with baseline models in terms of AUC and ACC matrices to validate the effectiveness of the NACD model. Second, we compare the performance of our selection strategies with the Random strategy and EM strategy using the informativeness metric. The experimental results demonstrate that our strategy outperforms other selection strategies. Third, we discuss the performance of our strategies compared to other strategies using the representativeness metric. Next, we present the visualization of the recommendation process of the KG-EIR strategy, EM strategy, and Random Strategy using heatmaps to highlight the excellent performance of the KG-EIR strategy. Finally, we explore the key components of the KG-EIR method further on the Eedi dataset.

\subsection{Datasets Descriptions}

We use two datasets in this paper: the Assistment (ASSIST) dataset and the Eedi dataset. The Assistment dataset is generated by collecting information from the Assistment Online Tutoring Systems. It is an open-source dataset for researchers to perform cognitive diagnosis tasks. The experiments in this paper are conducted on the problem bodies of this dataset.

The Eedi dataset is obtained from the NeuralIPS platform, which collected 233K records from 2064 students. Each student participated in an average of 112 workouts. For this study, we use problems 3 and 4 from the NeuralIPS dataset to compare the performance of our model.

The statistical information of the Assistment and Eedi datasets is shown in Table ~\ref{e_statistic}.

\begin{table}[!htbp] 
\centering
\caption{Statistics of the Assistment and Eedi datasets.\label{e_statistic}}
\begin{tabular}{c|c|c}
\toprule
\textbf{Statistic}	& \textbf{ASSIST} &\textbf{Eedi} \\
\midrule 
     Number of records & 267415 & 233000 \\ 
     Number of students & 2493 & 2064 \\ 
     Number of questions & 17671 & 948   \\ 
     Avg record/student & 107.2 & 112.8 \\ 
\bottomrule
\end{tabular}
\end{table}
To evaluate the KG-EIR method, we test it based on four standard cognitive diagnosis models: IRT, MIRT, NCDM, and KaNCDM. The details of these models are as follows:

\begin{enumerate}
	\item IRT \cite{embretson2013item}: This is the most popular CDM in computerized adaptive learning. IRT and conventional approaches focus on developing and applying multi-item scales to assess "latent variables" (hypothetical constructs).
	\item MIRT \cite{ackerman2003using}: MIRT is a multidimensional extension of IRT that demonstrates its potential for estimating several characteristics of ability. The IRT-based methods have also been expanded to accommodate MIRT.
	\item NCDM \cite{wang2020neural}: This cognitive diagnosis model is the most standard model in the educational data mining field. The NCDM model employs neural networks to learn the complex relationships of exercises in order to produce accurate and understandable diagnosis results.
	\item KaNCDM \cite{wang2022neuralcd}: This framework is further developed based on NCDM to estimate the current knowledge state of students. The KaNCDM improves upon NCDM in terms of feasibility, generality, and extensibility to make predictions. Extensibility is further discussed from two aspects: content-based extension and knowledge-association-based extension.
\end{enumerate}

We also apply two selection strategies to compare the performance of our selection strategy. The details of these selection strategies are as follows:

\begin{enumerate}
	\item Random strategy (RM): This strategy serves as the baseline for the selection strategies. It randomly selects questions from the question set without considering the overall performance.
	\item Expectimax strategy \cite{piech2015deep} (EM): Expectimax is a tree-based, brute-force MDP search algorithm that determines the expected utility of each action. It assumes that the agent will always choose the option that maximizes utility and that the environment will generate a subsequent state using a stochastic process after an action has been taken.
\end{enumerate}

\subsection{Framework Setting}

The framework settings for the KG-EIR framework are described in this part, as illustrated in Table~\ref{setting}.

\begin{table}[!htbp] 
\centering
\caption{The framework setting for the KG-EIR framework.\label{setting}}
\begin{tabular}{c|c|c}
\toprule
\textbf{}	& \textbf{ASSIST} &\textbf{Eedi} \\
\midrule 
     Attention embed size & 200 & 200 \\ 
     top-K & 5 & 5 \\ 
     Dropout rate & 0.2 & 0.2  \\ 
     Learning rate & 0.002 & 0.002 \\ 
     Number of epochs & 100 & 100\\
     $\alpha_1$ & 0.7 & 0.7\\
     $\alpha_2$ & 0.15 & 0.15\\
     $\alpha_3$ & 0.15 & 0.15\\
\bottomrule
\end{tabular}
\end{table}

\subsection{Results and Discussion}
In this paper, we evaluate the prediction task of the cognitive diagnosis model based on whether an exercise was successfully answered in the next interaction. We use the Area Under Curve (AUC) and Accuracy (ACC) metrics to measure students' performance in making predictions. A higher AUC or ACC value indicates better cognitive diagnosis performance, while a value of 0.5 suggests random selection. The cross-entropy loss function is used.

A binary value represents the effectiveness of exercise recommendation. we measure the performance of our selection strategy with the metric of Informativeness Metric and Coverage Metric. The Informativeness Metric(Inf(s)) is used for measuring the informativeness of the selection strategy in the exercise recommendation. The AUC metric is adopted to indicate the informativeness of the selection strategy referring to the following formula:
\begin{equation}
    Inf(s) = AUC({M(e_i,q_j|\Theta) | e_i \in E, q_j \in Q}) \label{i_auc}
\end{equation}
The Coverage metric(Cov(s) is accepted to measure the representativeness of the selection strategy. The Cov(s) is computed based on the percentage of knowledge concepts covered by the strategy-selected questions.

\begin{equation}
    Cov(s) = \frac{1}{|K|}\sum_{k\in K}1 [k\in Q_c] \label{i_q}
\end{equation}

\subsubsection{Student Prediction Performance}
Table \ref{q_res} presents a comparison of the results of baseline models with the Neural Attentive Cognitive Diagnosis (NACD) model. The NACD model outperforms all baseline models in terms of AUC and ACC on both the ASSIST and Eedi datasets. The MIRT model demonstrates better performance than the IRT model by extending it into the multidimensional space to predict students' performance. The NCDM model utilizes neural networks to further improve predictions and estimate the current state of students. The KaNCDM model enhances the NCDM model in terms of feasibility, generality, and extensibility, resulting in improved performance. The NACD-FE and NACD-FS models analyze the importance of exercise and student factors in predicting student performance. The results indicate that the student factor is more influential than the exercise factor, and the NACD model, which comprehensively considers both factors, outperforms all models.

\begin{table}[!htbp]
\centering
\caption{Comparison of results of baseline models with the Neural Attentive Cognitive Diagnosis model (NACD). The NACD outperforms all baseline models in terms of AUC and ACC.\label{q_res}}
\begin{tabular}{c|c|c|c|c}
\toprule
\multicolumn{1}{c}{\multirow{2}{*}{}}&\multicolumn{2}{c}{\textbf{ASSIST}}&\multicolumn{2}{c}{\textbf{Eedi}}\\
\midrule
& AUC & ACC  & AUC & ACC  \\
    IRT& 0.652 & 0.618 &0.687 & 0.634  \\
    MIRT& 0.666 & 0.645 &0.689 & 0.636 \\
    NCDM& 0.751 & 0.728 &0.709& 0.650\\
    KaNCDM & 0.768 & 0.728 & 0.747 & 0.683\\
    NACD-FE& 0.769 & 0.733 &0.748& 0.670 \\
    NACD-FS & 0.656 & 0.636  & 0.704 & 0.655 \\
     \textbf{NACD}& \textbf{0.772} &  \textbf{0.735} & \textbf{0.751} & \textbf{0.687}  \\ 
\bottomrule
\end{tabular}
\end{table}

\subsubsection{Informativeness Comparison}
\begin{figure}[h]
	\caption{Informativeness Comparison with AUC metric} 
	\label{infor_res}
	\centering
	\subfigure[The selection strategies performance on ASSIST datasets in timestamp 10 and timestamp 20]{
		\begin{minipage}{7cm} 
                        \includegraphics[width=\textwidth]{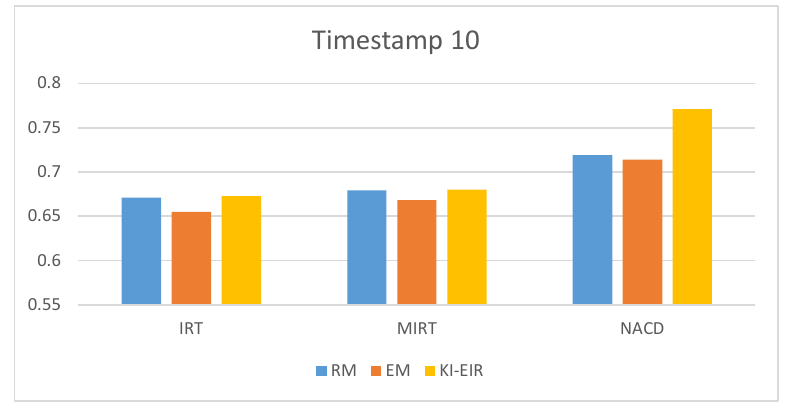} \\
		\end{minipage}
        \begin{minipage}{7cm}
			\includegraphics[width=\textwidth]{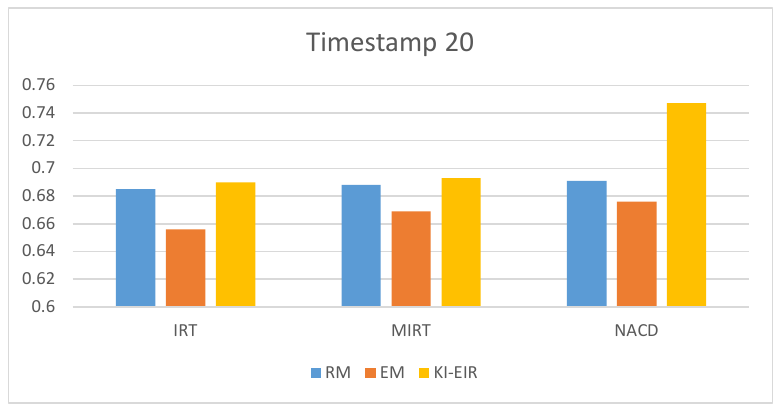} \\
			
		\end{minipage}
	}
	\subfigure[The selection strategies performance on Eedi datasets in timestamp 10 and timestamp 20]{
		\begin{minipage}{7cm}
			\includegraphics[width=\textwidth]{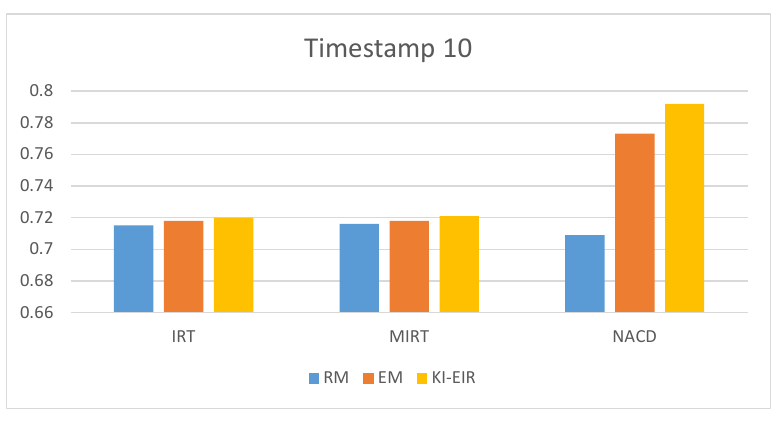} \\
			
		\end{minipage}
        \begin{minipage}{7cm}
			\includegraphics[width=\textwidth]{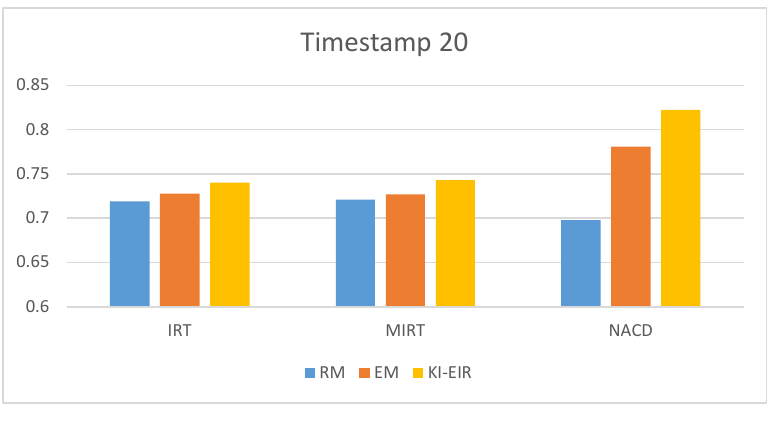} \\
			
		\end{minipage}
	}
\end{figure}

In this part, we compare the informativeness performance of different models and strategies using the AUC metric (Equation \ref{i_auc}). Figure \ref{infor_res} presents the results at the middle (step t = 10) and final (step t = 20) stages of the tests. We compare the KG-EIR strategy with the EM strategy, Random strategy, and the baseline models: NACD, IRT, and MIRT.

The random strategy performs the worst among all strategies on the Eedi dataset and provides the baseline accuracy for the experiment. The EM strategy, which utilizes a Markov Decision Process, outperforms the Random strategy on the Eedi dataset by considering the impact of interactions between exercises and students. However, the EM strategy performs worse than the IRT model on the ASSIST dataset due to the large number of exercises, which leads to inaccurate predictions when each exercise is treated as a state. The KG-EIR strategy, which incorporates exercise and skill features from the knowledge graph, outperforms all models on both datasets, indicating its effectiveness in achieving the informativeness goal. The KG-EIR strategy enhances the recommendation system by making the process more flexible without requiring modifications to the general methodology.

\subsubsection{Representativeness Comparison}

\begin{figure}[!htbp]
	\caption{Representativeness Comparison with Coverage metric.} 
	\label{PAAbefore}
	\centering
	\subfigure[IRT on Eedi and Assistment]{
		\begin{minipage}{7cm}
                        \includegraphics[width=\textwidth]{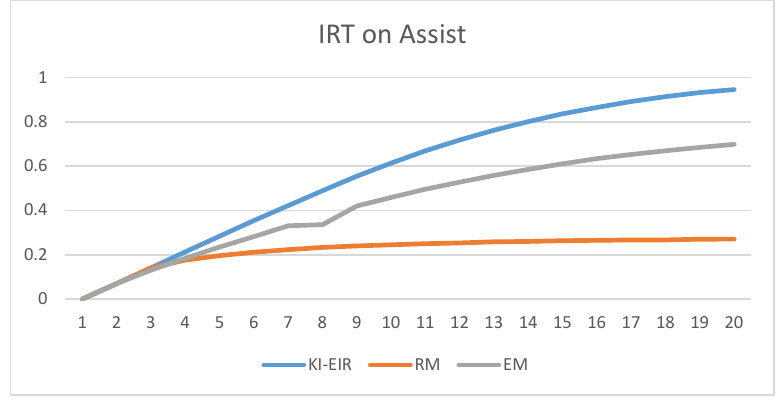} \\
		\end{minipage}
        \begin{minipage}{7cm}
			\includegraphics[width=\textwidth]{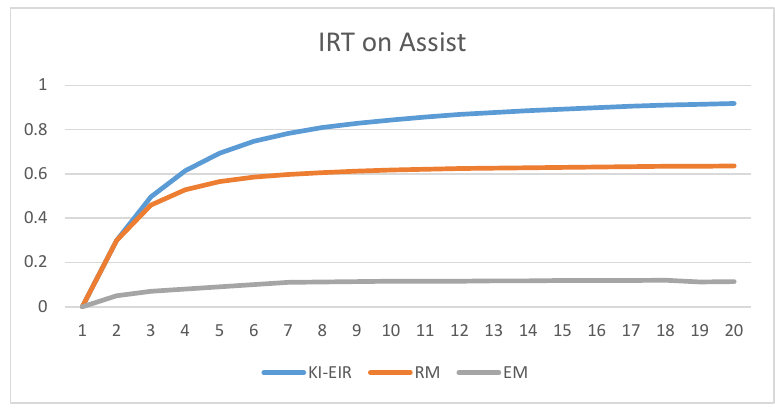} \\
			
		\end{minipage}
	}
	\subfigure[MIRT on Eedi and Assistment]{
		\begin{minipage}{7cm}
			\includegraphics[width=\textwidth]{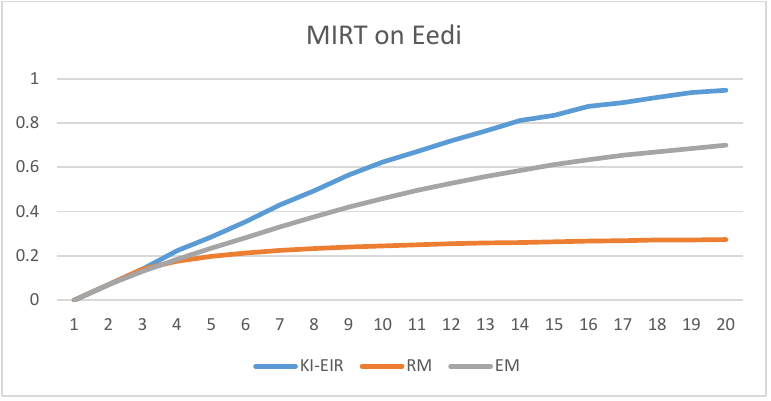} \\
			
		\end{minipage}
        \begin{minipage}{7cm}
			\includegraphics[width=\textwidth]{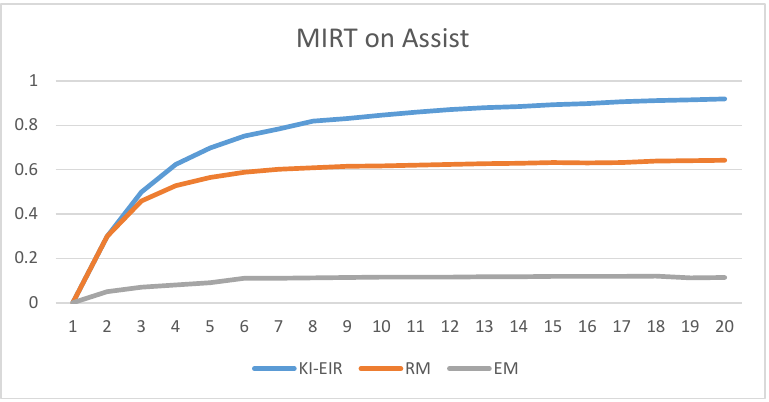} \\
			
		\end{minipage}
	}
 \subfigure[NACD on Eedi and Assistment]{
		\begin{minipage}{7cm}
			\includegraphics[width=\textwidth]{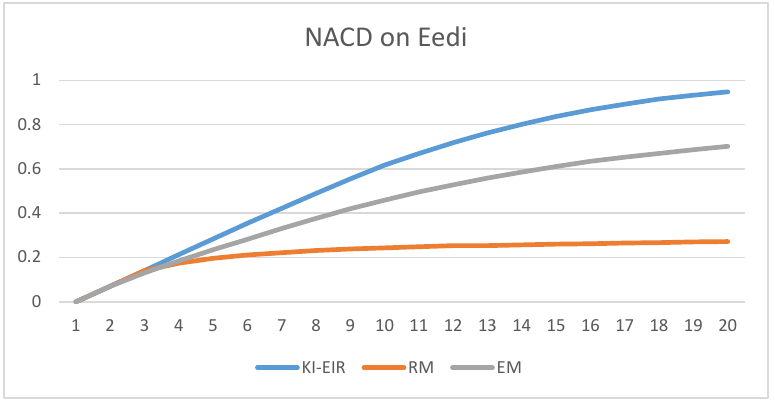} \\
			
		\end{minipage}
        \begin{minipage}{7cm}
			\includegraphics[width=\textwidth]{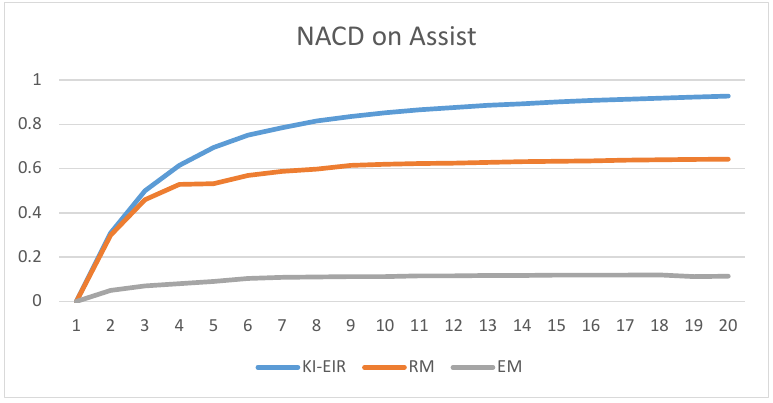} \\
			
		\end{minipage}
	}
\end{figure}

The representativeness comparison focuses on the coverage metric. The KG-EIR strategy outperforms other selection strategies on both datasets due to its incorporation of exercise features, skill features, and the knowledge graph. As shown in Figure \ref{PAAbefore}, the EM strategy performs better than the Random strategy because it considers the impact of behavior when selecting exercises on the Eedi dataset. However, the EM strategy performs worse than the Random strategy on the ASSIST dataset due to the large number of exercises, which leads to inaccurate predictions. The KG-EIR strategy, which incorporates exercise features, skill features, and the knowledge graph, achieves the best performance among all strategies. The KG-EIR strategy shows significant improvements in coverage metric, reaching close to 1 on both datasets. It consistently outperforms other strategies at all testing stages, demonstrating its effectiveness and flexibility in recommending exercises. The KG-EIR strategy enhances the recommendation system by providing informative and representative exercises based on the performance of the KG-EIR method.
 
\subsubsection{Visualization of Selection Strategies}

In this section, we validate the performance of the selection strategies, including the KG-EIR strategy, EM strategy, and Random Strategy, in recommending exercises to improve student performance on the Eedi dataset. Heatmaps are used to visualize the evolution of student performance, as measured by the AUC metric.

\begin{figure}[!htbp]
    \centering
    \includegraphics[width=\textwidth]{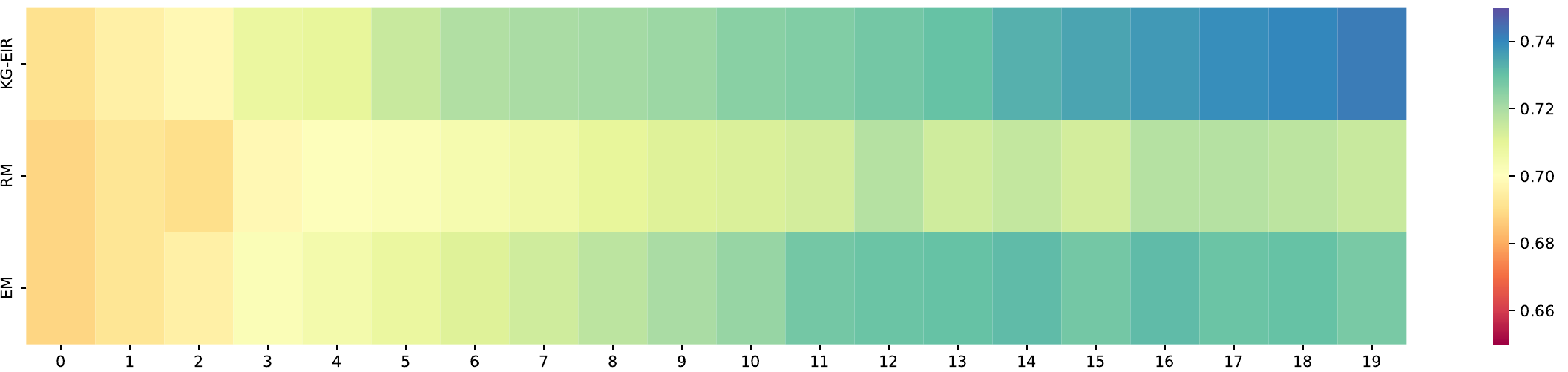}
    \caption{Heatmaps illustrating the performance of selection strategies on the Eedi dataset.}
    \label{h_Eedi}
\end{figure}

The heatmaps in Figure \ref{h_Eedi} depict the differences in performance based on different selection strategies (KG-EIR, EM, and Random) by observing the color change. The vertical dimension represents the selection strategies (KG-EIR, EM, and Random), while the horizontal dimension represents the different testing phases from 0 to 19. The color of the heatmap represents the performance of students when recommended with appropriate exercises, with stronger colors indicating a greater impact of the selection strategies.

According to Figure \ref{h_Eedi}, the Random strategy performs worse than the other selection strategies. The EM strategy, which considers the impact of behavior, outperforms the Random strategy in all testing phases on both datasets. However, the EM strategy does not consider exercise and skill features. The KG-EIR strategy, on the other hand, incorporates exercise and skill features and provides specific goals of informativeness and representativeness. As a result, the KG-EIR strategy achieves the best student performance among the selection strategies.

Overall, the results and discussions demonstrate the effectiveness and superiority of the proposed KG-EIR strategy in cognitive diagnosis and exercise recommendation. The KG-EIR strategy outperforms baseline models and other selection strategies in terms of informativeness and representativeness. It leverages exercise and skill features, along with the knowledge graph, to provide accurate and informative exercise recommendations for students. The visualization of the selection strategies further supports the outstanding performance of the KG-EIR strategy. These findings contribute to the improvement of cognitive diagnosis models and the enhancement of recommendation systems in educational settings.

\subsubsection{Ablation Experiments}

This section aims to identify the key components of the KG-EIR model through a series of ablation experiments. Four variations of the KG-EIR model are considered, each incorporating one or more components. Specifically, "IF," "ER," and "KI" indicate the removal of the informativeness component, exercise representativeness component, and knowledge importance component, respectively. "IF+KI" or "ER+KI" indicates the removal of the informativeness component and knowledge importance component or exercise representativeness component and knowledge importance component, respectively.

The conclusions drawn from the experiments are as follows. Firstly, the individual components of informativeness, exercise representativeness, and knowledge importance do not yield satisfactory outcomes when used alone. The performance gradually improves as more components are incorporated into the KG-EIR method. Secondly, the removal of the exercise representativeness component (ER) results in the most significant drop in AUC, decreasing to 71.8\%. Therefore, the knowledge importance component (ER) is more important than the informativeness component (IF) according to the experimental results in this section. Thirdly, since the knowledge importance component (KI) provides the skill weight for ER, when the ER component is removed, the KI component is also removed. Consequently, the inclusion of the KI component leads to greater performance improvement compared to the IF component.

Table \ref{e_ab} presents the results of the ablation study of the KG-EIR model based on the IRT model on the Eedi and Assist datasets.

\begin{table}[!htbp] 
\centering
\caption{Ablation study of the KG-EIR model based on the IRT model on two datasets.\label{e_ab}}
\begin{tabular}{c|c|c}
\toprule
\textbf{Method}	& \textbf{Eedi} & \textbf{Assist}\\
\midrule
Informativeness component(IF) & 0.719 & 0.670  \\ 
     Exercise Representativeness component(ER) & 0.718 & 0.672 \\ 
     Informativeness component(IF) + Knowledge Importance component(ER)& 0.720 & 0.676 \\ 
     Exercise Representativeness component(ER) + Knowledge Importance component(KI) & 0.721& 0.675  \\ 
      KG-EIR & 0.724& 0.679 \\
\bottomrule
\end{tabular}
\end{table}

The results in Table \ref{e_ab} demonstrate that the KG-EIR model outperforms the ablated versions in terms of AUC on both the Eedi and Assist datasets. The inclusion of all components in the KG-EIR model leads to the best performance. The ablation study confirms the importance of the informativeness, exercise representativeness, and knowledge importance components in the KG-EIR model, with their combined effect resulting in improved performance.

\section{Conclusion}\label{sec13}

In this paper, we proposed a comprehensive framework, the KG-EIR (Knowledge Graph Enhanced Exercise Item Recommendation) model, to address the challenge of providing informative and representative exercises in cognitive diagnosis tasks. The KG-EIR model consists of four key components: informativeness, exercise representation, knowledge importance, and exercise representativeness.

The informativeness component estimates the informativeness of each exercise and selects questions with high informativeness from the untested question set to the candidate question set. The exercise representation component utilizes the Graph Convolutional Network (GCN) model and two types of relation attention mechanisms to generate skill embeddings and exercise embeddings. The knowledge importance component applies the knowledge points extraction path algorithm and knowledge importance weighted algorithm to calculate the skill importance weight. Finally, the exercise representativeness algorithm combines the skill importance weight, exercise weight, knowledge coverage, response matrix, and dissimilarity matrix to select questions from the candidate question set into the tested question set with high representativeness. The NACD model is then employed to accurately estimate the state of students based on the selected exercises.

The KG-EIR model demonstrates promising results in improving cognitive diagnosis and exercise recommendation in educational settings. By leveraging the power of knowledge graphs and incorporating multiple components, our framework provides accurate and informative exercise recommendations for students, thereby enhancing their learning experience and academic performance. The proposed framework opens up new avenues for research and development in the field of educational data mining and cognitive diagnosis.

In future research, we suggest exploring the application of reinforcement learning techniques such as Deep Q-Network (DQN) to further improve the selection of exercises with high representativeness and informativeness. This approach can help reduce the time required for the selection phase and enhance the efficiency and effectiveness of the cognitive diagnosis process.

\backmatter

\bmhead{Acknowledgments}

The research work in this paper were supported by the National Natural Science Foundation of China (No. 62177022, 61901165, 61501199), AI and Faculty Empowerment Pilot Project (No. CCNUAI\&FE2022-03-01), Collaborative Innovation Center for Informatization and Balanced Development of K-12 Education by MOE and Hubei Province (No. xtzd2021-005), and Natural Science Foundation of Hubei Province (No. 2022CFA007).

\section*{Declarations}

The authors declare no conflict of interest.

\bibliography{sn-bibliography}

\end{document}